\documentstyle[12pt]{article}                    %%%
\newfont{\ffont}{msym10}                        %%
\newcommand{\beq}{\begin{equation}}             %%
\newcommand{\eeq}{\end{equation}}               %%
\newcommand{\bqry}{\begin{eqnarray}}            %%
\newcommand{\eqry}{\end{eqnarray}}              %%
\newcommand{\bqryn}{\begin{eqnarray*}}          %%
\newcommand{\eqryn}{\end{eqnarray*}}            %%
\newcommand{\preprint}[1]{\begin{table}[t]      %%
            \begin{flushright}                  %%
            \begin{large}{#1}\end{large}        %%
            \end{flushright}                    %%
            \end{table}}                        %%
\newcommand{\PD}[2]                             %%
    {\frac{\partial^{#2}}{\partial #1^{#2}}}    %%
\begin{document}
\preprint{ }
\title{Mass Spectrum of a Baryon Octet is Linear}
\author{\\ L. Burakovsky\thanks{Bitnet: BURAKOV@QCD.LANL.GOV} \
\\  \\  Theoretical Division, T-8 \\  Los Alamos National  
Laboratory \\ Los
Alamos NM 87545, USA \\  \\  and  \\  \\
L.P. Horwitz\thanks{Bitnet: HORWITZ@SNS.IAS.EDU. On sabbatical leave from
School of Physics and Astronomy, Tel Aviv University, Ramat Aviv, Israel.
Also at Department of Physics, Bar-Ilan University, Ramat-Gan,  
Israel  } \
\\  \\ School of Natural Sciences \\ Institute for Advanced Study  
\\ Princeton
NJ 08540, USA \\}
%  \\ and \\  \\ W.C. Schieve\thanks{Bitnet: WCS@MAIL.UTEXAS.EDU}\
%\\  \\ Ilya Prigogine Center \\ for Studies in Statistical Mechanics \\
%University of Texas at Austin \\ Austin TX 78712, USA \\}
\date{ }
\maketitle
\begin{abstract}
It is argued that the mass spectrum of a baryon octet is linear,  
consistent
with a Gell-Mann--Okubo type relation for the squared masses of the  
baryons
obtained previously by Bal\'{a}zs and Nicolescu. The mass spectrum of a 
baryon decuplet is briefly discussed.
\end{abstract}
\bigskip
{\it Key words:} hadronic resonance spectrum, Gell-Mann--Okubo

PACS: 12.40.Ee, 12.40.Yx, 14.20.-c
\bigskip
\section*{  }
The hadronic mass spectrum is an essential ingredient in theoretical 
investigations of the physics of strong interactions. It is well  
known that
the correct thermodynamic description of hot hadronic matter requires 
consideration of higher mass excited states, the resonances, whose
contribution becomes essential at temperatures $\sim O(100$ MeV)
\cite{Shu,Leut}. The method for taking into account these
resonances was suggested by Belenky and Landau \cite{BL} as considering 
unstable particles on an equal footing with the stable ones in the
thermodynamic quantities; e.g., the formulas for the pressure and energy 
density in a resonance gas read\footnote{For simplicity, we neglect the 
chemical potential and approximate the particle statistics by the
Maxwell-Boltzmann one.}
\beq
p=\sum _ip_i=\sum _ig_i\frac{m_i^2T^2}{2\pi  
^2}K_2\left(\frac{m_i}{T}\right),
\eeq
\beq
\rho =\sum _i\rho _i,\;\;\;\rho _i=T\frac{dp_i}{dT}-p_i,
\eeq
where $g_i$ are the corresponding degeneracies ($J$ and $I$ are spin and
isospin, respectively), $$g_i=\frac{\pi ^4}{90}\times \left[  
\begin{array}{ll}
(2J_i+1)(2I_i+1) & {\rm for\;non-strange\;mesons} \\
4(2J_i+1) & {\rm for\;strange}\;(K)\;{\rm mesons} \\
2(2J_i+1)(2I_i+1)\times 7/8 & {\rm for\;baryons}
\end{array} \right. $$
These expressions may be rewritten with the help of a {\it  
resonance spectrum,}
\beq
p=\int _{m_1}^{m_2}dm\;\tau (m)p(m),\;\;\;p(m)\equiv  
\frac{m^2T^2}{2\pi ^2}
K_2\left(\frac{m}{T}\right),
\eeq
\beq
\rho =\int _{m_1}^{m_2}dm\;\tau (m)\rho (m),\;\;\;\rho (m)\equiv
T\frac{dp(m)}{dT}-p(m),
\eeq
normalized as
\beq
\int _{m_1}^{m_2}dm\;\tau (m)=\sum _ig_i,
\eeq
where $m_1$ and $m_2$ are the masses of the lightest and heaviest  
species,
respectively, entering the formulas (1),(2).

In both the statistical bootstrap model \cite{Hag,Fra} and the dual  
resonance
model \cite{FV}, a resonance spectrum takes on the form
\beq
\tau (m)\sim m^a\;e^{m/T_0},
\eeq
where $a$ and $T_0$ are constants. The treatment of a hadronic  
resonance gas
by means of the spectrum (6) leads to a singularity in the thermodynamic 
functions at $T=T_0$ \cite{Hag,Fra} and, in particular, to an  
infinite number
of the effective degrees of freedom in the hadron phase, thus hindering a
transition to the quark-gluon phase. Moreover, as shown by Fowler  
and Weiner
\cite{FW}, an exponential mass spectrum of the form (6) is  
incompatible with
the existence of the quark-gluon phase: in order that a phase  
transition from
the hadron phase to the quark-gluon phase be possible, the hadronic  
spectrum
cannot grow with $m$ faster than a power.

In our previous work \cite{spectrum} we considered a model for a  
transition
from a phase of strongly interacting hadron constituents, described by a 
manifestly covariant relativistic statistical mechanics which  
turned out to be
a reliable framework in the description of realistic physical systems 
\cite{mancov}, to the hadron phase described by a resonance  
spectrum, Eqs.
(3),(4). An example of such a transition may be a relativistic high  
temperature
Bose-Einstein condensation studied by the authors in ref.  
\cite{cond}, which
corresponds, in the way suggested by Haber and Weldon \cite{HW}, to 
spontaneous flavor symmetry breakdown, $SU(3)_F\rightarrow SU(2)_I\times 
U(1)_Y,$ upon which hadronic multiplets are formed, with the masses  
obeying
the Gell-Mann--Okubo formulas \cite{GMO}
\beq
m^\ell =a+bY+c\left[ \frac{Y^2}{4}-I(I+1)\right];
\eeq
here $I$ and $Y$ are the isospin and hypercharge, respectively,  
$\ell $ is 2
for mesons and 1 for baryons, and $a,b,c$ are independent of $I$  
and $Y$ but,
in general, depend on $(p,q),$ where $(p,q)$ is any irreducible  
representation
of $SU(3).$ Then only the assumption on the overall degeneracy  
being conserved
during the transition is required to lead to the unique form of a  
resonance
spectrum in the hadron phase:
\beq
\tau (m)=Cm,\;\;\;C={\rm const}.
\eeq
Zhirov and Shuryak \cite{ZS} have found the same result on  
phenomenological
grounds. As shown in ref. \cite{ZS}, the spectrum (8), used in the  
formulas (3),(4) (with the upper limit of integration infinity), leads to
the equation of state $p,\rho \sim T^6,$ $p=\rho /5,$ called by  
Shuryak the
``realistic'' equation of state for hot hadronic matter \cite{Shu},  
which has
some experimental support. Zhirov and Shuryak \cite{ZS} have calculated 
the velocity of sound, $c_s^2\equiv dp/d\rho =c_s^2(T),$ with $p$  
and $\rho $
defined in Eqs. (1),(2), and found that $c_s^2(T)$ at first  
increases with $T$
very quickly and then saturates at the value of $c_s^2\simeq 1/3$  
if only the
pions are taken into account, and at $c_s^2\simeq 1/5$ if  
resonances up to
$M\sim 1.7$ GeV are included.

In order to understand why a linear spectrum is the actual spectrum of a 
hadronic multiplet, one may restrict himself to one family of the Regge 
trajectories (e.g., those on which the members of the baryon  
$J^P=\frac{1}{2}^{
+},\frac{3}{2}^{-},\frac{5}{2}^{+},\ldots $ octets lie), and  
calculate the
sum in Eq. (1) in two different ways:

1. Fix spin and calculate the sum in Eq. (1) for an individual  
multiplet, by
the introduction of the multiplet spectrum, $\tau (m),$ as in Eq.  
(3); then
with a spin degeneracy, $2J_i+1\simeq 2J_i\simeq 2\alpha  
^{'}m_i^2,$ involved,
$\alpha ^{'}\cong 0.84$ GeV$^{-2}$ being a universal Regge slope,  
the sum (1)
will reduce to Eq. (3) with
\cite{spectrum}
\beq
\tau ^{'}(m)\sim m^2\;\tau (m),
\eeq
where $\tau ^{'}(m)$ is the mass spectrum of a family of the Regge
trajectories.

2. Fix isospin and calculate the sum (1) for an individual
trajectory, then multiply the result by the number of the  
trajectories for an
individual multiplet (e.g., 8 for a baryon octet). Since the  
squared masses of
the states lying on a trajectory, are $m_0^2,$ $m_0^2+1/\alpha  
^{'},$ $m_0^2+
2/\alpha ^{'},\;\ldots ,$ the sum (1) may be calculated directly,  
using the
Euler-Maclaurin summation formula
\beq
\sum _{n=n_1}^{n_2}f_n=\int _{n_1}^{n_2}dn\;f(n)+\frac{1}{2}\left[f(n_1)+
f(n_2)\right]+({\rm derivative\;\;terms}),
\eeq
resulting in Eq. (3) with \cite{spectrum}
\beq
\tau ^{'}(m)\sim m^3.
\eeq
It then follows from (9),(11) that
\beq
\tau (m)\sim m,
\eeq
i.e., the mass spectrum of an individual hadronic multiplet is linear.

We have checked the coincidence of the results given by the linear  
spectrum
(8) with those obtained directly from Eq. (1) for the actual  
hadronic species
with the corresponding degeneracies, for all well-established hadronic 
multiplets, both mesonic and baryonic, and found it excellent  
\cite{spectrum}.
Shown are typical figures of ref. \cite{spectrum} in which the  
results given by
both, Eq. (1), and Eq. (3) with a linear spectrum, are  
compared,\footnote{
Instead of a direct comparison of Eqs. (1) and (3), we compared the  
expressions
$p/p_{SB}$ for both cases, where $p_{SB}\equiv \sum _ig_i\pi  
^2/90\;T^4,$
i.e., $p_{SB}$ is the pressure in an ultrarelativistic gas with  
$g=\sum _ig_i$
degrees of freedom.} for the following three well-established  
baryon octets:

$J^P=\frac{1}{2}^{+},$ $\;N(939),$ $\Lambda (1116),$ $\Sigma (1190),$ 
$\Xi (1320)$

$J^P=\frac{3}{2}^{-},$ $N(1520),$ $\Lambda (1690),$ $\Sigma (1670),$ 
$\Xi (1823)$

$J^P=\frac{5}{2}^{+},$ $N(1680),$ $\Lambda (1820),$ $\Sigma (1915),$ 
$\Xi (2030)$ \\ Thus, the theoretical implication that a linear  
spectrum is
the actual spectrum in the description of individual hadronic  
multiplets, is
consistent with experiment as well. In our recent papers  
\cite{su4,linear} we
have shown that the linear spectrum of an individual meson nonet is  
consistent
with the Gell-Mann--Okubo mass formula (as follows from (7))
\beq
m_1^2+3m_8^2=4m_{1/2}^2
\eeq
(in fact, this formula may be derived with the help of a linear spectrum 
\cite{su4}), and leads to an extra relation for the masses of the  
isoscalar
states, $m_{0^{'}}$ and $m_{0^{''}},$ (of which $0^{'}$ belongs to  
a mostly
octet),
\beq
m_{0^{'}}^2+m_{0^{''}}^2=m_0^2+m_8^2=2m_{1/2}^2,
\eeq
with $m_1,m_{1/2},m_8,m_0$ being the masses of the isovector,  
isospinor, and
isoscalar octet and singlet states, respectively, which for an  
almost ideally
mixed nonet reduces to \cite{su4,linear}
\beq
m_{0^{''}}^2\simeq m_1^2,\;\;\;m_{0^{'}}^2\simeq 2m_{1/2}^2-m_1^2.
\eeq
The relation (14) was checked in ref. \cite{linear} and shown to  
hold with
an accuracy of up to $\sim $3\% for all well-established nonets. In ref. 
\cite{su4} we have generalized a linear spectrum to the case of  
four quark
flavors and derived the corresponding Gell-Mann--Okubo mass formula  
for an
$SU(4)$ meson hexadecuplet, in good agreement with the experimentally 
established masses of the charmed mesons. In ref. \cite{enigmas} we have 
applied a linear spectrum to the problem of establishing the correct 
$q\bar{q}$ assignment for the problematic meson nonets, like the scalar, 
axial-vector and tensor ones, and separating out non-$q\bar{q}$ mesons.

By applying the arguments of refs. \cite{su4,linear}, one may show  
that the
linear spectrum of a baryon octet leads to a Gell-Mann--Okubo type  
relation
for the squared masses of the baryons. Indeed, for a baryon octet,  
one has 8
isospin states: $2N,$ $1\Lambda ,$ $3\Sigma $ and $2\Xi ,$ the  
$\Lambda $ and
$\Sigma $ states being mass degenerate on a naive quark model  
level, since
both are composed of two light $u$- and $d$-quarks and one heavier  
$s$-quark:
$m_\Lambda =m_\Sigma \equiv m_{\Sigma ^{'}}.$ The average mass  
squared of an
octet should coincide with that calculated with the help of a  
linear spectrum
\cite{su4,linear}; hence
\beq
\frac{2m_N^2+4m_{\Sigma ^{'}}^2+2m_\Xi ^2}{8}=\frac{m_N^2+m_\Xi ^2}{2},
\eeq
and therefore, $m_N^2+m_\Xi ^2=2m_{\Sigma ^{'}}^2,$ or
\beq
m_\Xi ^2-m_{\Sigma ^{'}}^2=m_{\Sigma ^{'}}^2-m_N^2.
\eeq
Let us check the relation (17) for the three well-established  
baryon octets
indicated above. We shall take $m_{\Sigma ^{'}}=(m_\Lambda  
+m_\Sigma )/2,$ with
$m_\Lambda $ and $m_\Sigma $ being the masses of the physical  
$\Lambda $ and
$\Sigma $ states, respectively.

1. $J^P=\frac{1}{2}^{+}$ octet. In this case $m_{\Sigma ^{'}}=1150$  
MeV, and
one has 0.42 GeV$^2$ on the l.h.s. of Eq. (17) vs. 0.44 GeV$^2$ on  
the r.h.s.

2. $J^P=\frac{3}{2}^{-}$ octet. Here $m_{\Sigma ^{'}}=1680$ MeV, and
one has 0.50 GeV$^2$ on the l.h.s. of Eq. (17) vs. 0.51 GeV$^2$ on  
the r.h.s.

3. $J^P=\frac{1}{2}^{+}$ octet. In this case $m_{\Sigma ^{'}}=1865$  
MeV, and
one has 0.64 GeV$^2$ on the l.h.s. of (17) vs. 0.655 GeV$^2$ on the  
r.h.s.

One sees that for the well-established baryon octets, the relation  
(17) holds
with a high accuracy, as for the standard (linear in mass)  
Gell-Mann--Okubo
formula (as follows from (7))
\beq
\frac{m_N+m_\Xi }{2}=\frac{3m_\Lambda +m_\Sigma }{4},
\eeq
which for these octets may be shown to hold with an accuracy of up  
to $\sim $
1\%.

The formula (17), together with the relation (as follows from (15))
\beq
m_{0^{'}}^2-m_{1/2}^2=m_{1/2}^2-m_{1^{'}}^2,\;\;\;m_{1^{'}}\equiv  
(m_1+m_{0^{
''}})/2,
\eeq
suggest the empirical rule for the splitting of the squared masses  
of the
hadrons containing zero, one, and two $s$- and/or $\bar{s}$-quarks, first
arrived at by Bal\'{a}zs and Nicolescu using two different  
approaches to the
confinement region of hadronic physics described in refs. \cite{BN,Bal}:
\footnote{The relation (19) holds for a close-to-ideally mixed meson
nonet, where $0^{'}\simeq s\bar{s}$ is composed almost of the  
$s$-quark and its
antiquark, and $0^{''}\simeq (u\bar{u}+d\bar{d})/\sqrt{2}$ is  
composed almost
of two light quark and antiquark and is mass degenerate with  
$1=(u\bar{u},
(u\bar{u}-d\bar{d})/\sqrt{2},d\bar{d}).$}  
$$m_{2s}^2-m_{1s}^2=m_{1s}^2-m_{0s}^
2.$$ If one assumes that this relation may be generalized to  
include also the
hadron containing three $s$- or $\bar{s}$-quarks, as follows:
\beq
m_{3s}^2-m_{2s}^2=m_{2s}^2-m_{1s}^2=m_{1s}^2-m_{0s}^2,
\eeq
one may apply it to a baryon decuplet. The only decuplet  
well-established
experimentally is $J^P=\frac{3}{2}^{+},$ $\Delta (1232),$ $\Sigma  
(1385),$
$\Xi (1534),$ $\Omega (1672).$ For this decuplet, Eq. (20) gives

$$m_{3s}^2-m_{2s}^2=m_\Omega ^2-m_\Xi ^2=0.44\;{\rm GeV}^2,$$

$$m_{2s}^2-m_{1s}^2=m_\Xi ^2-m_\Sigma ^2=0.435\;{\rm GeV}^2,$$

$$m_{2s}^2-m_{1s}^2=m_\Sigma ^2-m_\Delta ^2=0.40\;{\rm GeV}^2.$$ \\  
One sees
that the relation (20) holds for the baryon $J^P=\frac{3}{2}^{+}$  
decuplet with
an accuracy which is not lower than that of the standard  
Gell-Mann--Okubo mass
formula (as follows from (7))
\beq
m_\Omega -m_\Xi =m_\Xi -m_\Sigma =m_\Sigma -m_\Delta .
\eeq
We note that with the values $m_N=0.95$ GeV, $m_\Delta =1.22$ GeV,  
indicated in
\cite{BN}, in good agreement with experiment, the relations (17) and (20)
become almost exact for the baryon $J^P=\frac{1}{2}^{+}$ octet and  
$J^P=\frac{
3}{2}^{+}$ decuplet, respectively: now one has 0.42 GeV$^2$ on both  
sides of
Eq. (17), and $m_\Sigma ^2-m_\Delta ^2=0.43$ GeV$^2$ in Eq. (20).

Finally, let us briefly discuss the mass spectrum of a baryon decuplet. 
According to the arguments given above, the mass spectrum of a  
baryon decuplet
is linear, as well as for a meson nonet and a baryon octet. The  
average mass
squared of a decuplet calculated directly for 10 isospin states should, 
therefore, coincide with the value given by a linear spectrum, i.e.,
\beq
\frac{4m_\Delta ^2+3m_\Sigma ^2+2m_\Xi ^2+m_\Omega  
^2}{10}=\frac{m_\Delta ^2+
m_\Omega ^2}{2},
\eeq
leading to the relation
\beq
3m_\Sigma ^2+2m_\Xi ^2=m_\Delta ^2+4m_\Omega ^2,
\eeq
which, however, {\it does not} hold. The reason is that 10 isospin  
degrees of
freedom of the decuplet are distributed with a linear spectrum in a mass 
interval which is {\it broader} than $(m_\Delta,\;m_\Omega ).$  
Indeed, assuming
the validity of Eq. (20), i.e., $m_\Sigma ^2=m_\Delta ^2+a,$ $m_\Xi  
^2=m_\Delta
^2+2a,$ $m_\Omega ^2=m_\Delta ^2+3a,$ $a\simeq 0.43$ GeV$^2,$ it  
then follows
that
\beq
\langle m^2\rangle =\frac{4m_\Delta ^2+3m_\Sigma ^2+2m_\Xi  
^2+m_\Omega ^2}{10}=
m_\Delta ^2+a=m_\Sigma ^2;
\eeq
one therefore finds that the mass interval of the decuplet is  
$(\sqrt{m_\Delta
^2-a},\;m_\Omega )$, and the average mass squared given by a linear  
spectrum
for this interval coincides with (24):
\beq
\frac{1}{2}\left(m_\Delta ^2-a+m_\Delta ^2+3a\right)=m_\Delta  
^2+a=m_\Sigma ^2.
\eeq
Fig. 4 shows the ratio $p/p_{SB}$ calculated from both, Eq. (1)  
with the actual
baryon masses and degeneracies, and a linear spectrum with the mass  
interval of
the decuplet $(1.04\;{\rm GeV}\cong \sqrt{m_\Delta  
^2-a},\;1.672\;{\rm GeV}=
m_\Omega).$ One sees that the results coincide.

The generalization of a linear spectrum of a baryon multiplet to  
the case of
four quark flavors and the derivation of the corresponding  
Gell-Mann--Okubo
mass formula for an $SU(4)$ baryon 20-plet will be given in a separate
publication.

\section*{Acknowledgements}
One of us (L.B.) wish to thank E.V. Shuryak for very valuable  
discussions on
hadronic resonance spectrum.

\newpage
\centerline{FIGURE CAPTIONS}
\bigskip
\bigskip
\bigskip
\bigskip
\hfil\break
Fig. 1. Temperature dependence of the ratio $p/p_{SB}$ as  
calculated from:
a) Eq. (1), b) Eq. (3) with a linear spectrum, for the  
$J^P=\frac{1}{2}^{+}$
baryon octet, $N(939),$ $\Lambda (1116),$ $\Sigma (1190),$
$\Xi (1320).$\hfil\break
\hfil\break
\hfil\break
\hfil\break
Fig. 2. The same as Fig. 1 for the $J^P=\frac{3}{2}^{-}$ baryon octet,
$N(1520),$ $\Lambda (1690),$ $\Sigma (1670),$ $\Xi (1823).$\hfil\break
\hfil\break
\hfil\break
\hfil\break
Fig. 3. The same as Fig. 1 for the $J^P=\frac{5}{2}^{+}$ baryon octet,
$N(1680),$ $\Lambda (1820),$ $\Sigma (1915),$ $\Xi (2030).$\hfil\break
\hfil\break
\hfil\break
\hfil\break
Fig. 2. The same as Fig. 1 for the $J^P=\frac{3}{2}^{+}$ baryon decuplet,
$\Lambda (1232),$ $\Sigma (1385),$ $\Xi (1534),$ $\Omega (1672),$  
with the
mass interval for a linear spectrum $(1.04\;{\rm GeV},\;1.672\;{\rm  
GeV}).$
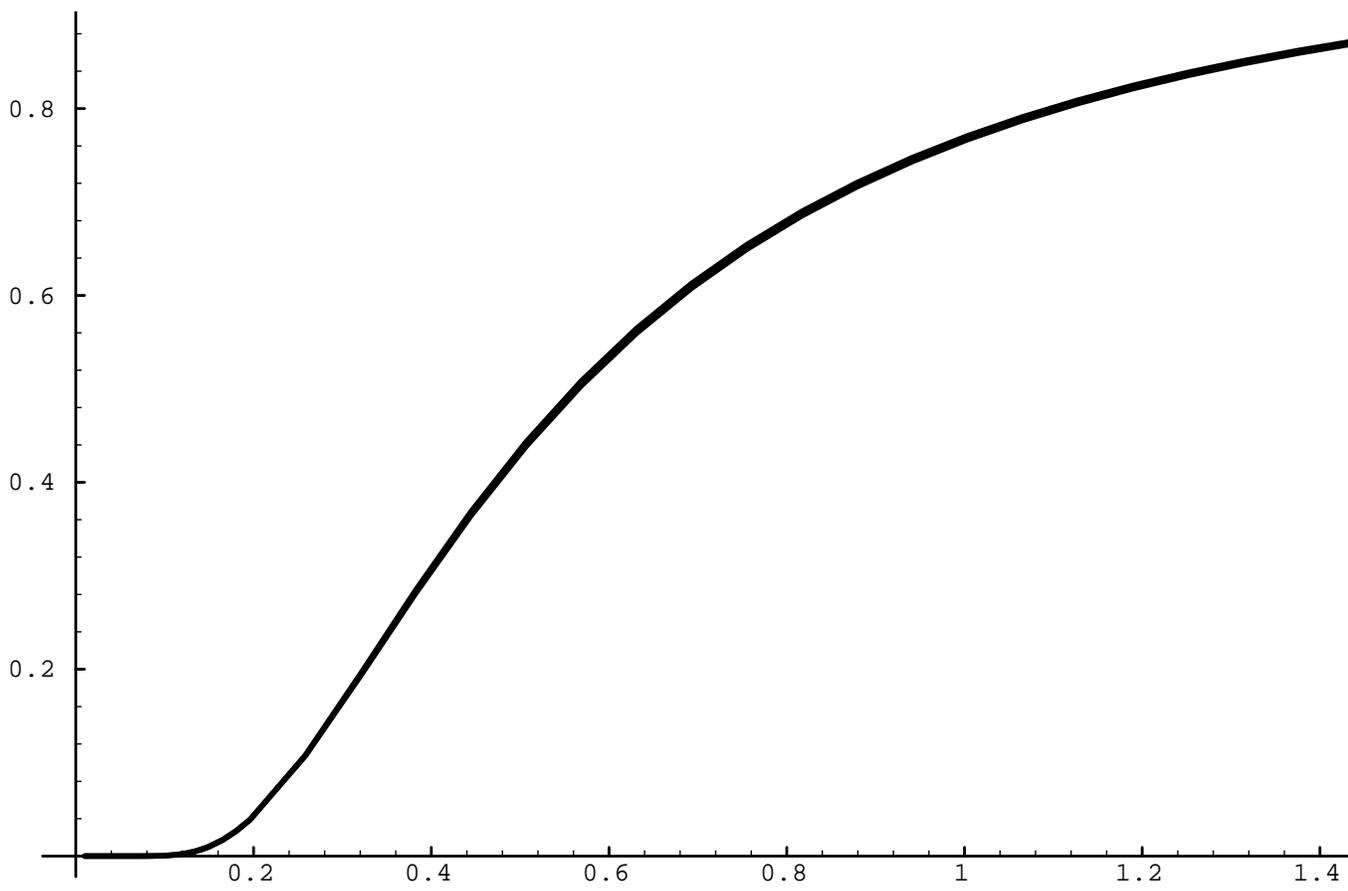
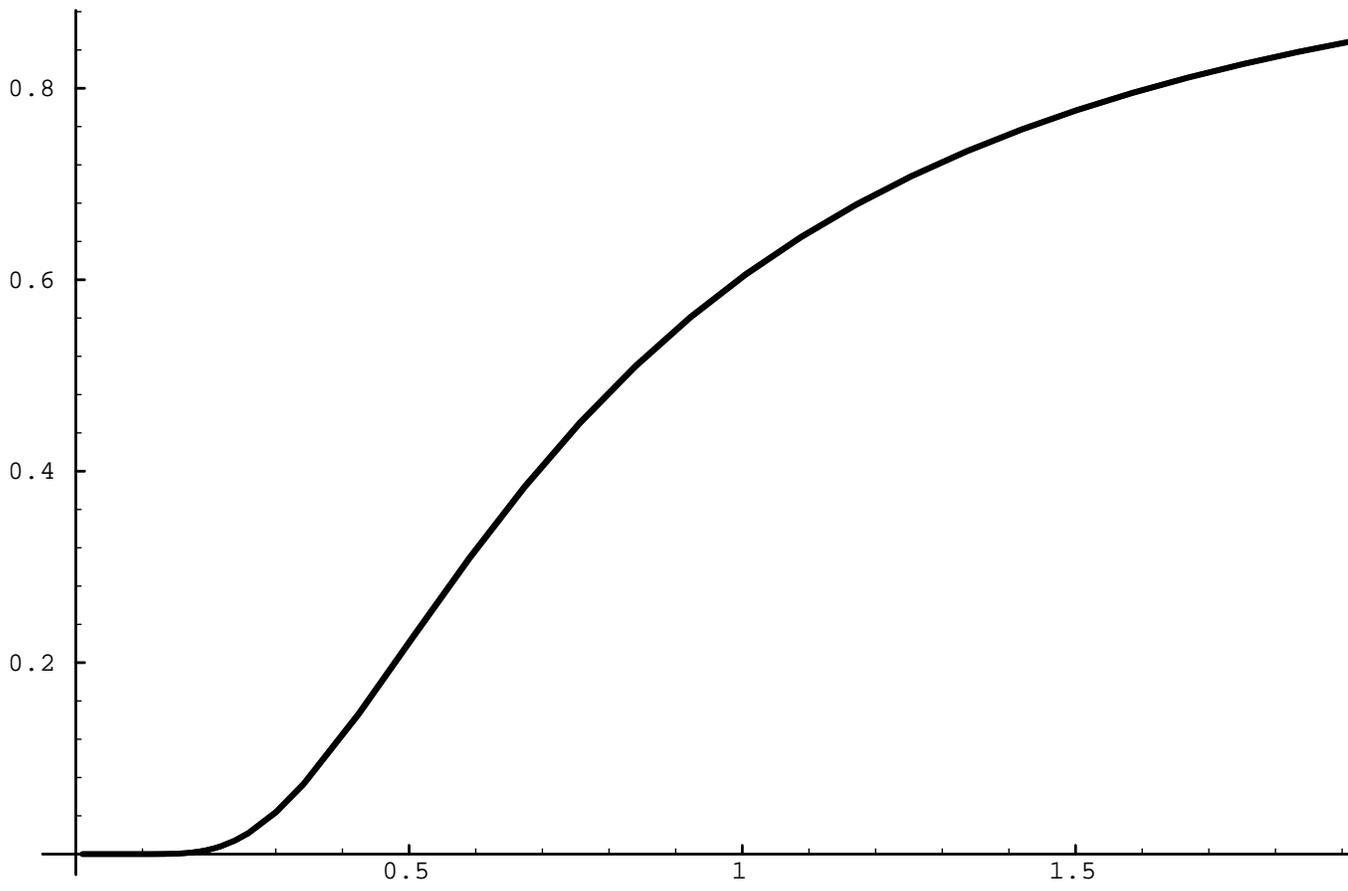
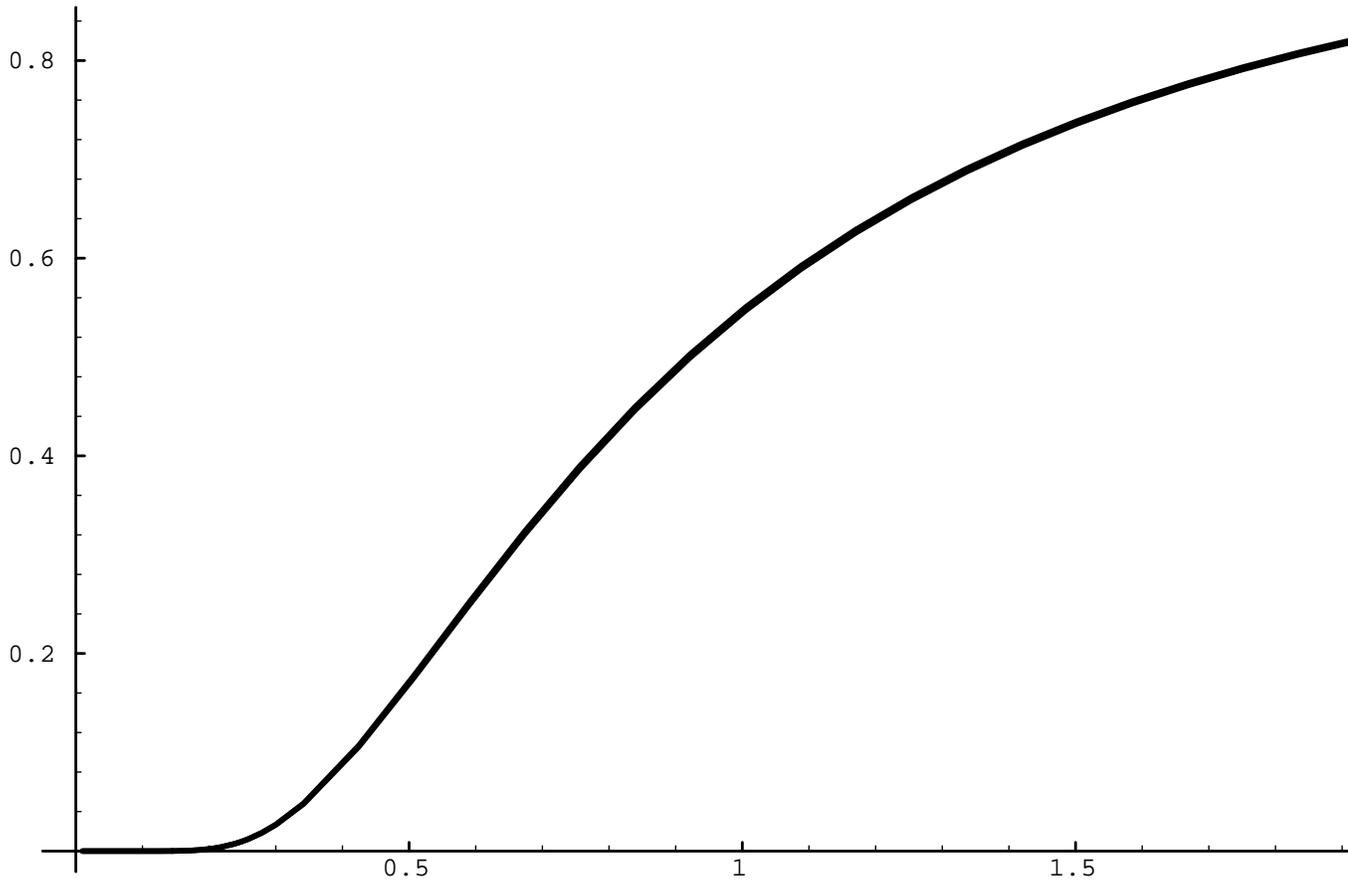
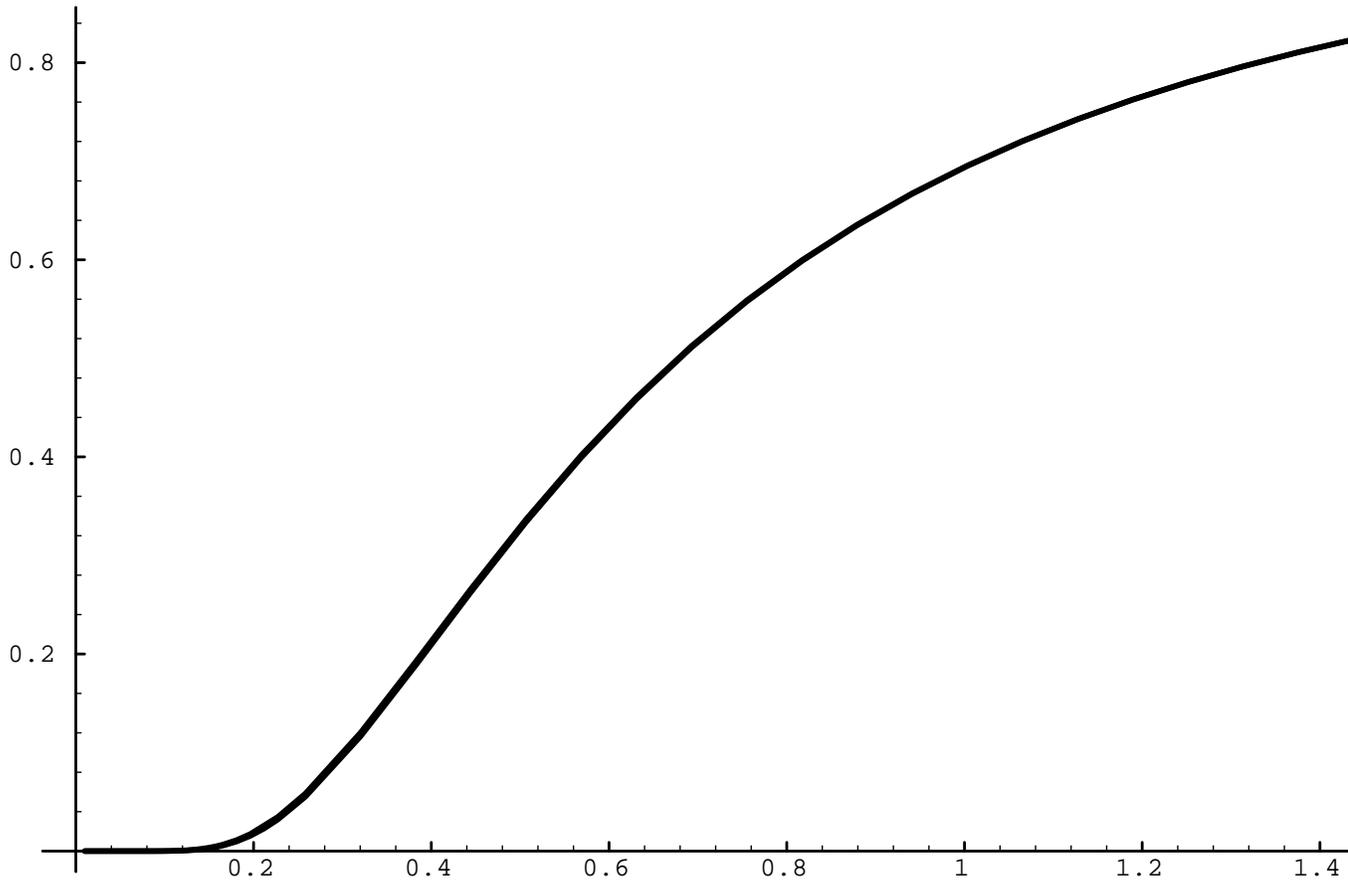
\end{document}